\documentclass[aps,prb]{revtex4} 
\usepackage{amsmath}
\usepackage{amsfonts}
\usepackage{graphicx}

\newcommand{\Sh}{Schr\"odinger{ }}

\newcommand{\mr}[1]{\mathrm{#1}}

\newcommand{\bs}[1]{\boldsymbol{#1}}

\newcommand{\fuve}[1]{\left(\bs{#1}\right)}
\newcommand{\fues}[1]{\left(#1\right)}

\newcommand{\yav}[1]{\left[#1\right]}

\newcommand{\llavl}[1]{\left\{#1\right.}

\newcommand{\abs}[1]{\left\vert#1\right\vert}

\newcommand{\prom}[1]{\left\langle #1 \right\rangle}

\newcommand{\Eq}[1]{Eq. (\ref{#1})}
\newcommand{\Eqs}[1]{Eqs. (\ref{#1})}

\begin{document}

\title[]{Hall field induced magnetoresistance oscillations
         of a two-dimensional electron system}

\author{A. Kunold} 
\email[ Email:]{akb@correo.azc.uam.mx} 
\affiliation{Departamento de Ciencias B\'asicas, Universidad Aut\'onoma Metropolitana-Azcapotzalco, Av. San Pablo 180,  M\'exico D. F. 02200, M\'exico}

\author{M. Torres}
\email[Email:]{torres@fisica.unam.mx} 
\affiliation{Instituto de F\'{\i}sica, Universidad Nacional Aut\'onoma de M\'exico, Apartado Postal 20-364,  M\'exico Distrito Federal 01000,  M\'exico}

\date{\today}

\begin{abstract}

We develop a model of  the nonlinear response to  a DC electrical 
current  of a two  dimensional electron system(2DES) placed on  a magnetic field. 
Based on   the exact solution of the  \Sh equation in arbitrarily strong  electric and  magnetic    fields, and  separating the  relative and guiding center coordinates,  a Kubo-like   formula for the current   is worked out   as a response to  the impurity scattering.  Self-consistent expressions determine the longitudinal and Hall components of the electric field in terms of the  DC current. 
The differential resistivity displays strong  Hall field-induced  oscillations,  in 
agreement with the   main features of the phenomenon   observed in recent experiments. 

\end{abstract}

\maketitle 

\section{Introduction}
Recently, non-equilibrium magnetotransport in high mobility two-dimensional
electron systems (2DES)  has acquired great experimental and theoretical interest.
Microwave-induced resistance oscillations (MIRO) attaining zero resistance states were discovered a few years ago \cite{Zudov0,Mani0,Zudov1} in 2DES  subjected to microwave irradiation and moderate
magnetic fields, corresponding to very high  
Landau levels (LL$s$).
The   phototoresistance exhibits   strong oscillations  periodic  in
$\epsilon=\omega/\omega_c$, where $\omega$ and $\omega_c$ are the  microwave
and cyclotron frequencies.
Our current understanding of this phenomenon  rest upon  models that  predict  the existence of negative-resistance states yielding an  instability that rapidly drive
the system into a zero resistance state \cite{Andreev1}.
Two distinct mechanism for the generation of negative resistance states are known,  one is based in the 
microwave induced impurity scattering \cite{Ryz,Durs,Lei0, Shi0, Vavi0, Torres0}, while the second  is linked to the modifications of the distribution produced by  the microwaves and  inelastic processes \cite{Dmi1,Dmi2,Ken}. 
Similar to MIRO are magnetoresistance  oscillations induced by the combined
effects of  microwave irradiation  and periodic potential modulation \cite{Dietel0,Torres1,Yuan0}.

More recently, Hall field-induced resistance oscillations (HIRO) have been observed
in high mobility samples in response  to a strong  DC electric current \cite{Yang0,Bykov0}.
The HIRO  oscillations are  periodic in inverse 
magnetic field,  with  the resistance maxima appearing at   integer values 
of the dimensionless parameter $\epsilon =  \omega_H /\omega_c $.
The Hall frequency $\omega_H = \gamma   \left( 2 \pi /n_e\right)^{1/2} J_x /e$, 
is associated  to  the energy $\hbar \omega_H = \gamma R_c \, E_y^{cl} $; where the
the classical Hall electric field is given by  $E_y^{cl} = B J_x/ e n_e$,  $R_c$ is the cyclotron radius of the electrons at 
the Fermi level, and  $\gamma \sim 1.63 \,\, to \,\, 2.18$.  
These results were confirmed in the recent experiment by   Zhang $et \, al.$  \cite{Zhang0},
with a determination of the parameter $\gamma \sim 1.9$. 
Additionally in this work another notable nonlinear effect was found: in the
regime of separated LL$s$ a relative weak DC current  induces a dramatic reduction of the resistivity.
Although MIRO and HIRO are basically different phenomena, both rely on the
commensurability of the cyclotron frequency with a characteristic parameter,
$\omega$ and $\omega_H$ respectively.
The study of HIRO permits to analyze  resistivity as a function of $\epsilon$
when both $\omega_c$ and $\omega_H$ are varied over a wide 
range of frequencies.  Instead the MIRO studies  are carried out  performing 
 $\omega_c$-sweeps at fixed $\omega$, because the  experimental difficulty in implementing 
 $\omega$-sweeps.  Other interesting examples of nonlinear magnetotransport experiments
combine DC and AC excitations \cite{Bykov0,Zhang1}.

 Some theoretical studies of  the nonlinear transport properties under DC exitations have recently appeared.  
  In the work of Vavilov $et \, al.$  \cite{Vavi1} the nonlinear  magnetotransport effects are related 
  to  changes of the electron distribution function induced by the DC electric field. 
 On the other hand, the work of Lei \cite{Lei1}  the impurity scattering processes are analyzed  utilizing  a semiclassical method in  which the evolution of the CM velocity operator and relative electron energy are obtained from the Heisenberg operator equations, and  the CM electron  coordinate and   velocity are treated classically. 
The aim of this work is to develop a model to describe the nonlinear response to  a DC electrical 
current  of a 2DES placed on  a magnetic field.
Our model  is based on the exact
solution of the Shr\"odinger  in arbitrarily strong  magnetic  and electric fields. 
 It is convenient to separate the electron coordinate in  its relative $\bs{R}$  and guiding center $\bs{X}$  coordinates. Although the relative and guiding center coordinates commute; the $R_x$ and $R_y$ relative coordinates satisfy a non-commutative algebra (and similarly for the $X$ and $Y$ projections of the relative 
 coordinates). These properties are exploited  to work out a Kubo-like formula  for the  conductivity 
  as a response to  randomly distributed impurities. Self-consistent expressions determine the longitudinal and Hall components of the electric field in terms of the  DC current  density  $J_x$.
  The Hall electric field $E_y$ is given by the dominant classical  Hall contribution   $E_y^{cl} = B J_x/ e n_e$,  plus an small quantum correction $\Delta E_y$.   The  differential  resistance displays  a strong oscillatory behavior  resulting from current  assisted electron scattering producing both   intra-LL and inter-LL  transitions. 
The results of the present work taken together with those of Lei \cite{Lei1} prove that the  properties  of HIRO oscillations in the  overlapping LL$s$  region  ($\epsilon > 1$), as well as the  strong resistance  reduction in the separated LL  regime  can be both  explained by a DC-current  induced impurity scattering mechanism.

\section{Model}

 The goal  of this work is to study the nonlinear  response to  a DC electrical 
current  density  ($J_x$)  of a  2DES  placed in  a magnetic field.   The electrons are subjected  both  to  a magnetic 
$  \bs B = \fues{0,0,B}$  and an   in-plane  electric  $\bs E=\fues{E_x,E_y,0}$ fields. Additionally  the effects of   impurity  scattering potential $V$ are  included; hence the dynamics is governed  by the Hamiltonian 
 
\begin{equation}
H = H_{\{B,E\}} +  V \, , 
\label{ham0:1}
\end{equation}
where 
\begin{equation}
H_{\{B,E\}}  = 
\frac{1}{2m^*} \bs \Pi^2+e\bs{E}\cdot \bs x \, , 
\label{ham0:2}
\end{equation}
includes the interaction with the magnetic and electric fields,  the covariant momentum is  $\bs \Pi = - i \hbar  \bs{\nabla}+e\bs{A}$. 
The impurity  potential is decomposed  in terms of its Fourier
components
\begin{equation}\label{imppot}
V\fues{\bs{r}}=\sum_i^{N_i}\int\frac{d^2q}{\fues{2\pi}^2}V\fues{q}
\exp\yav{i\bs{q}\cdot\fues{\bs{r}-\bs{r}_i}} \, , 
\end{equation}
where $\bs{r}_i$ is  the position of the $i$th impurity and $N_i$ is the number
of impurities.  The explicit form of $V\fues{q}$ depends on the nature of the 
scatterers. 
For  short-range
neutral  impurities   \cite{Torres0}:  $V\fues{q}= 2 \pi \hbar^2 \alpha /m^*$, where 
$\alpha $ is the scattering length, and 
the impurity density is related to the electron mobility according to the relation 
$ \alpha^2 n_{imp} = e/(4 \pi^2 \hbar \mu)$.  Instead,  in the case  of a $2D$ screened  Coulomb 
potential:  $V \fues{q}  = \frac{\pi\hbar^2 q_{TF}  }{m^* }  e^{-q d} / \left( q_{TF} + q \right) $,  where $d$ is
the thickness of the  doped layer and $ q_{TF} =  e^2 m^*  / \left( 2 \pi \epsilon_0 \epsilon_b \hbar^2 \right)$; in this case
the relation of the  impurity density to the electron mobility is approximated as  
 $  n_{imp} = 8 e (k_F d)^3 /(\pi \hbar \mu)$.

A planar electron performs cyclotron and drifting motion in magnetic and electric  fields.  It is then convenient 
to decomposed the electron coordinate $\bs r$ into the guiding center $\bs{X} = (X,Y)$, and the 
the relative coordinate  $ \bs{R} = (R_x,R_y)$, $i.e.$:   $\bs r= \bs X  + \bs R$, where 
$\bs R = ( -\Pi_y/eB \, , \, \Pi_x/eB)$.  The commutation relations are 

\begin{equation}\label{comm1}
\left[ R_x , R_y \right]= \frac{ i \hbar}{ e  B} \, ,   \hskip1.4cm \left[ X  ,  Y  \right]= - i l_ B^2 \, , 
 \hskip1.4cm \left[ X_i ,  R_j  \right]=  0  \, , 
\end{equation}
with $l_B^2= \hbar / e B$. With this  decomposition the  $X$ and $Y$  coordinates  become  noncommutative (similarly for $R_x$ and $R_y$); however   the guiding center and the relative coordinates are independent variables.

For an arbitrary orientation of the electric field  $ \bs E= E \fues{\cos \theta ,\sin \theta,0}$, 
the Hamiltonian $H_{\{B,E\}}$ can be exactly diagonalized  if the vector potential is selected in the
 gauge:  $ \bs A= B \fues{y\, \sin^2 \theta \, , \, -\, x \,\cos^2 \theta, \,0}$. The spectrum and eigenstates 
 are given by 
 \begin{eqnarray}\label{solshe}
 {\cal E} _{\{\mu,k\}}  &=&  \hbar \omega_c \left( \frac{1}{2} + \mu  \right) + e E l_B  k   -  \frac{(e E l_B)^2}{2 \hbar \omega_c}  \, , \nonumber \\
 \Psi_{\mu,k } &= &  \frac{1}{ \sqrt{2^\mu \mu ! \sqrt{\pi} l_B} } \,   \exp\{- i  k \,  \eta \}  \,  \exp\{- (\xi - \xi_c)^2/2 l_B^2 \}  \,  H_\mu \left[
  ( \xi - \xi_c)/l_B^2 \right]  \, , 
 \end{eqnarray}
 here  $ \xi = x \cos \theta + y \sin \theta$, and  $ \eta = -x \sin \theta + y \cos \theta $ are the longitudinal and transverse coordinates (with respect to the electric field); 
 $ \xi_c  = \frac{e l_B^2  E} {\hbar \omega_c } - l_B  k $,  and $H_\mu$  represent  the Hermite polynomials. The index $k$ is the eigenvalue of the transverse $ - X \sin \theta + Y \cos \theta$  center of guide coordinate.

We now turn to the calculation of the current density. The velocity of the center guide coordinates are obtained from
the Heisenberg operator equations using  the total Hamiltonian in   \Eq{ham0:1}: 
\begin{equation}
\dot X=\frac{i}{\hbar} \yav{H ,X}=  \frac{E_y}{B}  - \frac{l_B^2}{\hbar} \, 
 \frac{\partial V }{\partial y} \, , \hskip2.5cm 
\dot Y=\frac{i}{\hbar} \yav{H,Y}=-  \frac{E_x}{B}  +  \frac{l_B^2}{\hbar} \, 
\frac{\partial V }{\partial x} \, . 
\label{Heq}
\end{equation}
The  current density  is computed from the impurity and thermal average
$\prom{\bs{J}}=e \langle \mr{Tr}\yav{\rho\fues{t}\bs V} \rangle$ of the center of guide  velocity $\bs V = (\dot X, \dot Y ) $, 
weighted with the  density matrix  that satisfies the von Neumann  equation 
\begin{equation}\label{vNe}
i\frac{\partial}{\partial t}\rho=\yav{H_0+V,\rho} \, . 
\end{equation}

We shall now derive the Kubo-Greenwood  formula for the current within the framework of
the linear response theory with respect to the impurity potential. 
The electric field effects are exactly taken into account through the  wave function solution given in
  \Eq{solshe}.  The Hamiltonian is split into a unperturbed part $H_{\{B,E\}}$ and into the perturbation 
 $V(\bs{r}) \, \exp\left( - \delta \vert t \vert\right) $; notice that  we added to the impurity potential a term $ \exp\left( - \delta \vert t \vert\right) $,
 with  $\delta$ representing  the rate at which the perturbation is turned on and off.   The density 
 is similarly decomposed as $\rho = \rho_0 + \Delta \rho$, where the unperturbed density matrix takes the form $\rho_0 = \sum_\alpha f({\cal E}_\alpha) \vert \alpha \rangle \langle \alpha \vert$; where $f$ is the Fermi distribution function. Following  the usual procedure  of the linear response formalism, 
 the matrix elements of $  \Delta \rho$, in the base given by states in \Eq{solshe},  are worked out as

 \begin{equation}\label{melem}
  \langle \mu, k  \vert  \, \Delta \rho   \, \vert \nu, k^\prime   \rangle =   \langle \mu, k  \vert  \, V \, \vert \nu, k^\prime   \rangle 
  \left( f_{\nu, k^\prime} -  f_{\mu, k} \right)  \, \frac{1}{ {\cal E} _{\nu, k^\prime } - {\cal E} _{\mu, k } + i \delta } \, .
\end{equation}
Combining the previous equations, and assuming randomly distributed impurities, the average current density is explicitly computed;  as suggested by 
   \Eq{Heq}  the current is spplited in  drifting and  impurity  scattering contributions:
 
 \begin{equation}\label{curr1}
    J_i (\bs{E}) =   \epsilon_{ij} \,  \frac{e  n_e}{B} E_j  \, + \,      J_i^{(imp)} (\bs{E})  \, ,  \hskip1.5cm 
   J_i^{(imp)} (E_x,E_y)= \epsilon_{ij}\int \frac{d^2q}{(2\pi)^2} q_j F(\omega_{\bs{q}} )      \, , 
\end{equation}
here   the function $F$ is given by 
 
 \begin{equation}\label{funF}
 F( \omega_q ) = -  \frac{ e n_I }{\hbar} \abs{V\fuve{\bs{q}}}^2
\sum_{\mu,\nu}\abs{D_{\mu,\nu}\fues{\tilde q}}^2\yav{f\fues{E_\mu + \omega_q}-f\fues{E_\nu}}
\mr{Im}G\fues{E_\mu + \omega_q - E_\nu }\\
\end{equation} 
and is evaluated at arbitrary values of the electric field through the argument dependence according to 
\begin{equation}\label{wq}
\omega_q =    e l_B^2 \left( q_y E_x - q_x E_y\right) \, . 
\end{equation}
The matrix elements $D_{\mu,\nu}$ are given by
\begin{eqnarray}\label{matD}
D_{\mu,\nu} (\tilde q) =\exp\fues{-\frac{\abs{\tilde q}^2}{2}}\llavl{\begin{array}{ll}
\tilde q ^{\mu-\nu}\sqrt{\frac{\nu!}{\mu!}}L_\nu^{\mu-\nu}\fues{\abs{\tilde q }^2},&
\mu\ge\nu,\\
\fues{-{\tilde q}^*}^{\nu-\mu}\sqrt{\frac{\mu!}{\nu!}}L_\nu^{\nu-\mu}\fues{\abs{\tilde q}^2},&
\mu\le\nu,\\
\end{array}}
\end{eqnarray}
with  $\tilde q =(q_x-iq_y)/\sqrt{2}$ and $L_\nu^{\mu-\nu}\fues{\abs{\tilde q }^2}$ denotes
the associated Laguerre polynomial.

   A formal procedure to obtain the Green function    requires a self-consistent 
    calculation using the Dyson equation for the self-energy with the magnetic,  impurity, phonon, and other scattering effects included. 
     A  detailed calculation of  $Im \, G $  incorporating  all these elements is beyond the scope of the present  work. Instead we choose a gaussian-type expression for the  the density of states.  This expression can be justified   within a 
   self-consistent Born  calculation  that incorporates the   magnetic field and  disorder  effects 
   \cite{Ando0,Gerhardts0,Gerhardts1,Ando1},   hence  the  density of states for the $\mu$-Landau level   is represented  as 
 
\begin{equation}\label{dense}
Im \, G \fues{\omega_{\bs{q}}+E_\nu-E_\mu}
=\sqrt{\frac{\pi}{2\Gamma^2}}
\exp\yav{-\frac{\fues{\omega_{\bs{q}}+E_\nu-E_\mu}^2}{2\Gamma^2}},\quad\quad
\Gamma^2=\frac{2\beta \hbar^2\omega_c}{\tau_{tr}}.
\end{equation}
The parameter $\beta$ depends on the difference of the transport scattering
time $\tau_{tr}$ obtained from the mobility and the single particle scattering
time $\tau_s$.  
In the case of short-range scatterers  $\tau_{tr} = \tau_s$ and $\beta=1$. In the case of long-range
screened potential,  $\beta_\nu$  depends on the filling factor $\nu$,  $e.g.$ 
$\beta_{\nu=50} \approx 13.5$. \cite{Torres0}

The current in \Eq{curr1}   applies  in general in the non-linear transport  regimen, for  an arbitrary strength  of  the DC electric field.
The linear  response  regime  is recovered if 
the function $ F(\omega_q) $ is expanded to first order in   $E$  (the zero order term cancels because of the angular integration) to yield:   $ J_x  = \sigma_{xx} \, E_x $ and  $ J_y = \sigma_{yy} \,  E_y $,  where 
\begin{equation}\label{sigxx}
\sigma_{xx} = \sigma_{yy} =    e l_B^2  \int \,  \frac{d^2q}{(2\pi)^2} \, \,  q_y^2 \, \, \left(  \frac{\partial F }{\partial \omega_q} \right)_{\omega_q
= 0 } 
\end{equation}
and  $\sigma_{xy} = \sigma_{yx} = e n / B$.

Let us now consider the nonlinear transport regime.  In a typical experimental configuration the electric field 
is not explicitly controlled, instead  the longitudinal current $J_x$  is  fixed to a constant value, while  the transverse Hall current $J_y$ cancels. Consequently  \Eqs{curr1}
lead to the conditions

 \begin{align}\label{curr2}
       J_x   &= \hskip0.3cm  \frac{e n_e}{B} E_y   +      J_x^{(imp)} (E_x, E_y)   \,  ,     \nonumber \\
       0  &=  -\frac{e n_e}{B} E_x    +    J_y^{(imp)}  (E_x, E_y)   \, . 
\end{align}
 They represents two implicit equation for the unknown   $E_x$ and  $E_y$;
 the equations can  be solved following a self-consistent iteration.
 However, it is easily verified  that for the conditions that apply  in recent experiments,
 the solution of the previous equations simplify  assuming  that the 
 following conditions  $E_x \ll E_y$ and  $ J_x^{(imp)}     \ll    e n_e E_y /B $
 are  simultaneously satisfied.  It is then possible to explicitly solve  for 
  $E_x$  and $E_y$

 \begin{align}\label{fieldE1}
       E_y   & =  \frac{B}{e n_e } J_x   -       \frac{B}{e n_e }   J_x^{imp} (E_x, E_y)     \approx \frac{B}{e n_e } J_x  -  
      \frac{B}{e n_e }    E_x  \left( \frac{ \partial  J_x^{imp} (E_x, E_y)  }{\partial E_x } \right)_{(E_x = 0 , E_y = B J_x / e n_e )}  \, , 
     \\
       E_x  &  =  \frac{B}{e n_e } J_y^{imp}  (E_x , E_y )    \approx \frac{B}{e n_e } J_y^{imp}  (E_x= 0 , E_y  = B J_x / e n_e)  \, .
       \label{fieldE2}
\end{align}
In order to derive the previous results  we used the fact  that
 $ J_x^{imp} (E_x=0, E_y) $  cancels because of the angular integration. 
\Eq{fieldE1}  shows that the leading contribution to the Hall electric field is given by the classical 
 result  $   E_y^{cl} =  B J_x  / e n_e  $, however there is  a correction  $\Delta E_y$  given by the second term of the RHS. of \Eq{fieldE1}.  Utilizing  \Eqs{curr1}  the correction to the Hall electric  field can be worked out as

\begin{equation}\label{delE}
\Delta E_y  =    \frac{\hbar \, E_x}{e n_e} \int \,  \frac{d^2q}{(2\pi)^2} \, \,  q_y^2 \, \, \left(  \frac{\partial F }{\partial \omega_q} \right)_{\omega_q =  \omega_q^* }  \, , 
\end{equation}
where $E_x$ is determined by \Eq{fieldE2}  and  $  \omega_q^*$ is given by 

\begin{equation}\label{wqe}
\omega_q^* =  \frac{q_x  J_x }{e n_e } \, .
\end{equation}

\begin{figure} [hbt]
\begin{center}
\includegraphics[width=4.5in]{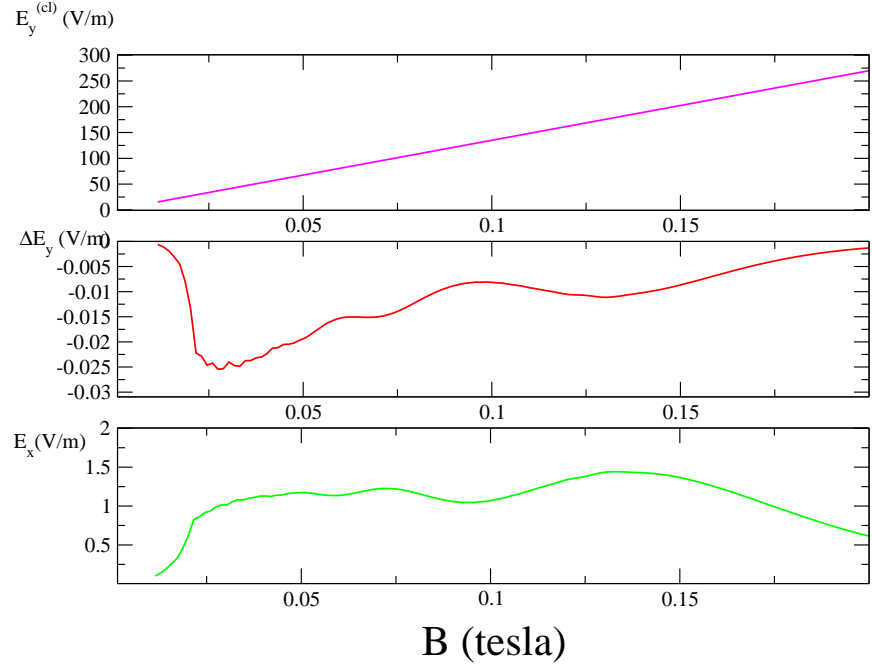}
\end{center}
\caption{(color online)
 Induced electric Hall field:  classical $E^{cl}_y \, (B)$ and  quantum correction $\Delta E_y \, (B)$,   and  longitudinal electric 
field $E_x \, (B)$) calculated from  Eqs. (\ref{fieldE1}) and (\ref{fieldE2}) at $J_x = 0.8 \, A/m$. The values of the other parameters are: $m^*=0.0635 m_e$,  $\mu=1.2\times 10^7 \, cm^2/Vs$, $n_e=3.7 \times 10^{11} cm^{-2}$, $T=1.5K$, and $\beta = 5$. }
\label{figure1}
\end{figure}

Figure \ref{figure1} shows  the electric fields: $E_y^{cl} =  B J_x / e n_e$,  $\Delta E_y$, and $E_x$ as a function of the magnetic  field for a fixed value of the longitudinal current. Notice that the conditions $\Delta E_y  \ll E_x \ll  E_y^{cl} $ hold in general,  validating the assumed approximation. In fact it is verified that the approximated solutions in Eqs. (\ref{fieldE1}) and (\ref{fieldE2}) coincide with  the self-consistent solutions that are obtained from \Eqs{curr2} with a  better that $1 \%$ precision.

Collecting the previous  results, it follows that the Hall resistivity is well approximated 
by the expression 

\begin{equation}\label{rhoxy}
\rho_{xy} = \frac{E_y}{J_x} =   \frac{B}{e n_e }   -  
      \frac{B E_x }{e n_e J_x }     \left( \frac{ \partial  J_x^{imp}   }{\partial E_x } \right)_{(E_x = 0 , E_y = B J_x / e n_e )}  \approx \frac{B}{e n_e } \, .
\end{equation}

Whereas the expression for  the non-linear longitudinal resistivity is given as 
\begin{equation}\label{rhoxx}
\rho_{xx} = \frac{E_x}{J_x} = -  \frac{B}{e n_e  J_x }  \,  \int \frac{d^2q}{(2\pi)^2} q_x F(\omega_q^* )  \, ,
\end{equation}

with $ \omega_q^*$  given in \Eq{wqe}. The differential resistivity  is calculated as $r_{xx} = \partial \left( J_x \rho_{xx} \right) / \partial J_x $
yielding 

\begin{equation}\label{rxx}
r_{xx} =   -  \frac{\hbar B}{\left( e n_e \right)^2 }     \int \frac{d^2q}{(2\pi)^2} q_x^2  \left(  \frac{\partial F }{\partial \omega_q} \right)_{\omega_q =  \omega_q^* }  \, .
\end{equation}

\Eqs{curr1}   for the nonlinear current together with  the definitions in Eqs. (\ref{funF}-\ref{dense})   constitute the central result of the paper.  They apply in general in  the nonlinear
regime in which both the longitudinal  and Hall electric fields are arbitrarily strong. However,  for the conditions that apply in 
 the experiments of current interest, it is reasonable to consider the $E_x$ weak limit. Then,  the Hall field  is accurately approximated by the classical result $E_y   = B J_x/ e n_e $, whereas   $\rho_{xx}$ and the  differential resistivity  are explicitly computed from \Eq{rhoxx} and  \Eq{rxx}  respectively.

In the work of  Zhang $et \, al.$  \cite{Zhang0}  the Hall   frequency is defined $\omega_H=  \gamma J_x \fues{2 \pi/e^2  n_e}^{1/2}$.
Here we assume that $\gamma = 2$, this   can be justified if we observe that the integral in \Eq{rhoxx} is evaluated in terms of the  variable  $\omega_q^* = q_x J_x / e n_e$ and  it is dominated by contributions  of exchanged  momentum  in the region 
$ q_x \approx 2 k_F$. Recalling that $k_F = \sqrt{2 \pi n_e} $, it yields 
   $\omega_H=   J_x \fues{8 \pi/e^2  n_e}^{1/2}$.  The 
dimensionless control  parameter is then  given by the ratio of the Hall to the cyclotron frequencies   

\begin{equation}\label{eps}
\epsilon = \frac{ \omega_H }{\omega_c } =  \frac{ 2 e E_y R_c }{\hbar \omega_c} \, ,  \hskip2.0cm R_c= \frac{v_F}{\omega_c}  \, , 
\end{equation}
and  it can be interpreted as the ratio of the work of the electric Hall field associated with the displacement of the guiding center of the 
cyclotron trajectory by $2 R_c$ to the Landau energy $\hbar \omega_c$.


\section{Results}
Results are presented for  a 2DES  $GaAs$ sample,  with  parameters corresponding to those
 reported in  recent experiment of  Zhang $et \, al.$  \cite{Zhang0}:
$m^*=0.0635 m_e$ , electron mobility $\mu=1.2\times 10^7 \, cm^2/Vs$  and  density $n_e=3.7 \times 10^{11} cm^{-2}$,  and  lattice temperature $T=1.5K$. 
For the impurities we consider an average of short- and long-range scatterers,  selecting the parameter $\beta$  
that appear in the   \Eq{dense}   as $\beta = 5.  $
\\

\begin{figure} [hbt]
\begin{center}
\includegraphics[width=4.5in]{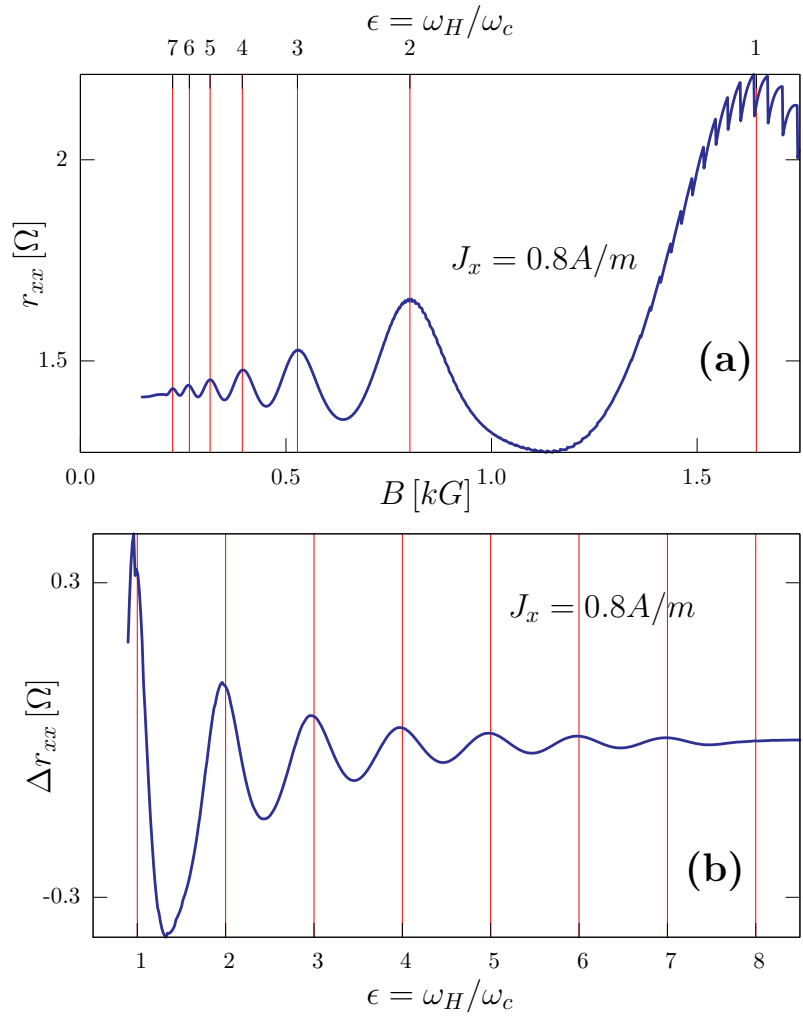}
\end{center}
\caption{(color online)
(a) Differential resistance $r_{xx} \, (B) $. (b) Correction to differential resistance  $\Delta r_{xx} = r_{xx} - r_{xx}(J_x = 0)$  versus $\epsilon =\omega_H/\omega_c$.  
 The  parameters 
  have the same values as in  Fig(\ref{figure1}).}
\label{figure2}
\end{figure}

Fig. \ref{figure2} (a)  shows the  differential resistance $r_{xx}$
as a function of the magnetic field $B$ for a fixed  current density $J_x = 0.8 A/m$  $(\omega_H /2 \pi \approx 65 \, GHz)$. We observe clear  differential magnetoresistance oscillations  mounted on an offset of $1.41 \Omega$ determined by the Drude contribution.
 At the top of this figure the values
of $\epsilon=\omega_H/\omega_c$  are displayed,  suggesting an  oscillations period 
$\Delta \epsilon \sim  1$.
To confirm these observations,  in 
 Fig. \ref{figure2} (b)   the Hal-field  induced  correction $\Delta r_{xx} = r_{xx} - r_{xx}(J_x = 0)$ is plotted as a function
of $\epsilon$.  Magnetoresistance oscillations are clearly observed  up to the seventh order. The first peak appears at $\epsilon \sim 0.95$, 
  for higher $\epsilon$ oscillations  the maxima occur  at  $\epsilon \approx  j$, with $j$ an integer; while  the   minima are  very close to
 $ \epsilon \approx j + 1/2$.
 These  results are  very similar to the experimental findings of Zhang $et \, al.$  \cite{Zhang0}, although the localization of maxima (minima) close to integer (half integer) are here  obtained 
when $\gamma = 2$, whereas in that work they correspond to the selection  $\gamma = 1.9$. 
   The amplitude of the differential resistance oscillations display a rapid decay  as the
magnetic field decreases. 
This  decay   can  be parametrised  by the Dingle factor $ \delta = exp \left( - \pi /\omega_c \tau_s \right) $, this allow us to get an estimate of the single particle scattering  time $\tau_s \approx 15 \, ps $, that compares  with the transport scattering time as $\tau_s / \tau_{tr}
\approx 4.8$ ; in good agreement with the selected value $\beta =  5$ .
\\

\begin{figure} [hbt]
\begin{center}
\includegraphics[width=4.5in]{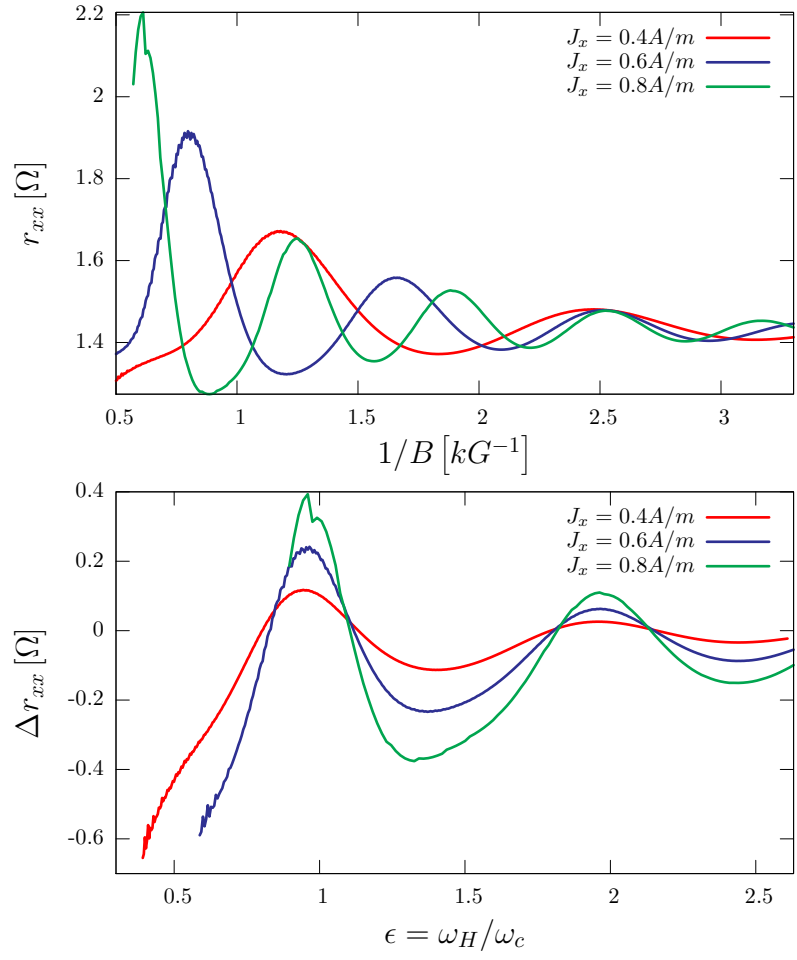}
\end{center}
\caption{(color online)
(a) Differential resistance $r_{xx}$ as a function  $1/B$ and (b) Correction to the differential
resistance $\Delta r_{xx} \, (\epsilon)$
for: $J_{x}=0.4 A/m$ (red), $J_{x}=0.6 A/m $ (blue) and $J_{x}=0.80 A/m$ (green, higher peak).}
\label{figure3}
\end{figure}

To   further support the previous  results,  we present  in 
Figure \ref{figure3} (a)  a plot of  $r_{xx}$  as a function
of  $1/B$ for various  values of the longitudinal current  density $J_{x}$. As
expected each curve shows a different period as a function of $1/B$, though the decay of the oscillation amplitudes are well described by the same Dingle factor $ \delta = exp \left( - \pi /\omega_c \tau_s \right) $,  with  $\tau_s \approx 15 \, ps $.  Notice that the  all the curves 
share a maximum at $B=1/2.5 KG$ given that $0.4 \,  A/ m$, $0.6 \, A/m$ and $0.8 \, A/m$ are
multiples of $0.2  \, A /m$.  Notwithstanding, when 
 $\Delta r_{xx} $  is plotted  as functions of $\epsilon$ in Fig. \ref{figure3} (b),   the positions of the maxima  (minima) of the  various curves  draw  close to integer (half integer)  values of $\epsilon$. 

\begin{figure} [hbt]
\begin{center}
\includegraphics[width=4.5in]{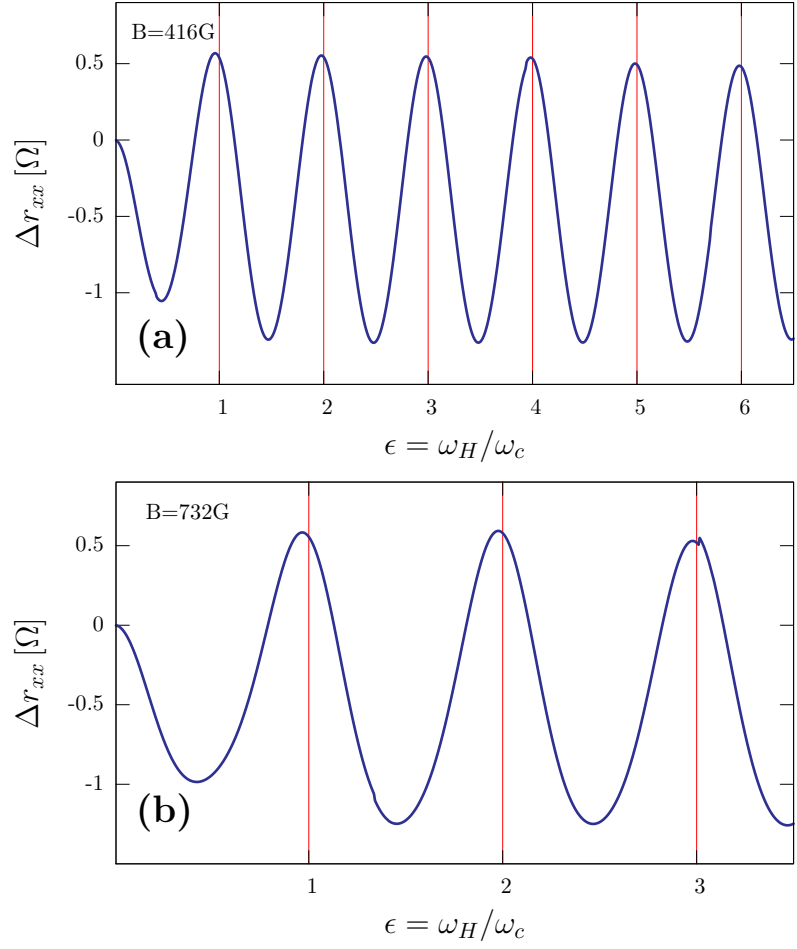}
\end{center}
\caption{(color online)
Correction to differential resistance $\Delta r_{xx} \, (\epsilon)$  obtained from $J_x$-sweeps at fixed values of   $B=416G$ (a)  and  $B=712G$ (b).}
\label{figure4}
\end{figure}

The oscillation waveform can be more clearly appreciated in graphs in which $\epsilon$ 
varies  performing current-sweepes  at fixed  $B$.  In   \ref{figure4} (a) and (b)  plots of 
 $\Delta r_{xx} \,  (\epsilon)$  are displayed for $B= 416 \, G$ and $B = 732 \, G$.
 We again observe that the first maxima appears at $\epsilon  < 1$, but as $\epsilon $ increases, 
 the position of the maxima (minima) are localized  very close to integer (semi-integer) values of 
 $\epsilon$. The   oscillation amplitude is  almost  constant,  because  
 the Dingle factor now remains constant as $\epsilon $ varies.

\begin{figure} [hbt]
\begin{center}
\includegraphics[width=4.5in]{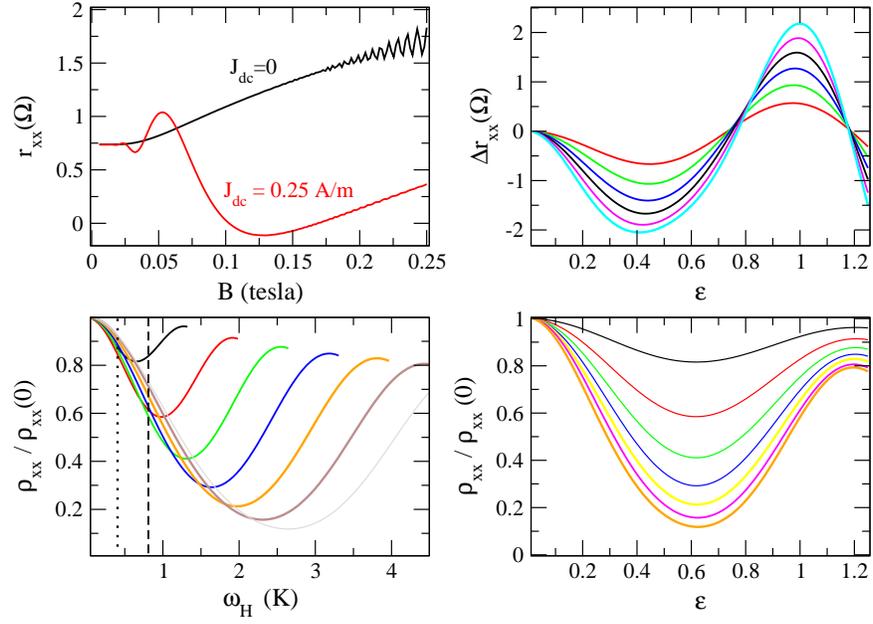}
\end{center}
\caption{(color online)
(a) Differential resistance $r_{xx} \, (B)$ for  $J_{x} = 0$ (black)      and  $J_{x} = 0.25 A/m$ (red).
(b) Correction to differential resistance $\Delta r_{xx} \, (\epsilon)$ obtained    varying $J_x$
at  fixed values of $B$, from   $B= 0.5\, kG$ (red)
to  $B= 1.75 \,  kG$ (light blue)   in $0.25 \, kG$  steps (top to botton  at $\epsilon=0.4$).
(c) Normalized longitudinal resistance  as a function of $\omega_H$ for the same selection of values of the magnetic fields
as in (b). The vertical line shows the value of $2 \, \Gamma$, \Eq{dense}, for $B=0.5 \, kG$ (dotted) and 
$B=1.75 \, kG$ (dashed) showing good agreement with the half-width of  the $J_x$  decay of  
$\rho_{xx}/\rho_{xx}(0)$.
(d) This figure  displays similar plots as a function of $\epsilon$.  
}
\label{figure5}
\end{figure}

We finally focus our attention to the regime of separated LL$s$, 
$\epsilon  < 1$. 
 Fig. \ref{figure5} (a)  displays  the differential resistivity as a function of the magnetic field for
both zero DC bias and a small current $J_x = 0.25 \, A/m$.   Notice that for the selected  small current ($J_x = 0.25 \, A/m$)  only one HIRO peak is resolved, however  in the region above above this  peak  ($\omega_H < \omega_c$) the resistivity is strongly  suppressed.  This suppression is in very good agreement with the experimental results of  Zhang $et \, al.$  \cite{Zhang0}, and   resembles the observed suppression of resistance observed for MIRO. The suppression is also observed in 
\ref{figure5} (b) where $J_x-$ sweeps are performed in order to plot $\Delta r_{xx} (\epsilon)$ 
at fixed values of $B$; several values of $B$ are selected, from top to bottom 
$B = 0.75,1.0,1.25,1.5,1.75,2.0  \, kG$. It is observed that the minima becomes deeper as $B$ is increased,  but in all the cases the  the maxima 
lies very  near $\epsilon = 1$.  The values of the width  $\Gamma$ (\Eq{dense})  are marked  
in the plot by the vertical dotted ($B= 0.75 \, kG$) and dashed  ($B= 2.0 \, kG$) lines. Considering 
the expression for the density of states in \Eq{dense}, we can interpret the decaying of $\rho_{xx}$ 
in the region   $\omega_H < \Gamma$ as a result of intra-Landau transitions;  whereas in the  region 
$ 2 \, \Gamma <  \omega_H $ the resistivity increases as a result of  the inter-Landau transitions. A similar behavior is observed when the resistivity is plotted as a function of $\epsilon $; however
now the minima of $\rho_{xx}$ is attained at $\epsilon = 0.6$ for all   values of  $B$. The minima 
of  $\rho_{xx}$ corresponds to the region $  2 \, \Gamma <  \omega_H < \omega_c - 2 \, \Gamma $ 
in which both intra-Landau and inter-Landau transitions are both suppressed.

\section{Conclusions}

In conclusion we presented a theory for the non-linear transport of a two 2DES placed in a magnetic field. The non linear response to a   DC current is incorporated by the exact solution of the 
\Sh equation including the effects of  arbitrarily strong  magnetic  and  in-plane electric fields. By means of the  non-conmuting relative and guiding center coordinates, a linear Kubo formula with respect to the impurity scattering is worked out. 
 The nonlinear expression for the electric  current  Eqs. (\ref{curr1}-\ref{wq})   constitute the central result of the paper.  They apply in general in  the nonlinear
regime in which both the longitudinal  and Hall electric fields are arbitrarily strong. However   in the experiments of current interest, it is reasonable to consider the $E_x$ weak limit in which  the Hall field  is accurately approximated by the classical result $E_y   = B J_x/ e n_e $, thus   $\rho_{xx}$ and the  differential resistivity  $r_{xx}$  are explicitly computed from \Eq{rhoxx} and  \Eq{rxx}  respectively.  
Our model is able to reproduce the most important features of
recent experiments\cite{Yang0,Bykov0,Zhang0}.  In the region    of separated LL$s$
the differential resistance as a functions of $\epsilon$ presents  strong oscillations with constant period ($\epsilon = 1$); the dominant mechanism being the  current-induced  impurity tunneling  between Landau levels.    In the region of of separated LL$s$ the  
dramatic reduction of the resistivity  at relative weak field is well reproduced;  the origin being 
related to the suppression of scattering within the LL$s$. 

\acknowledgments
We acknowledge  financial support  by
CONACyT  \texttt{G 32736-E},  and UNAM project No.  \texttt{IN113305}.

\bibliography{article}


\end{document}